%% file: SWrenDeepCoreNMONuPhys2015Proceedings.tex
\documentclass[12pt]{article}
\usepackage{graphicx}
\usepackage[]{units}

\def\Manchester{School of Physics and Astronomy, The University of Manchester\\
Oxford Road, Manchester, M13 9PL, United Kingdom}

\def\Title#1{\begin{center} {\Large #1 } \end{center}}
\def\Author#1{\begin{center}{ \sc #1} \end{center}}
\def\Address#1{\begin{center}{ \it #1} \end{center}}

\newenvironment{Abstract}{\begin{quotation}  }{\end{quotation}}
\newenvironment{Presented}{\begin{quotation} \begin{center} 
             PRESENTED AT\end{center}\bigskip 
      \begin{center}\begin{large}}{\end{large}\end{center} \end{quotation}}
\def\Acknowledgements{\bigskip  \bigskip \begin{center} \begin{large}
             \bf ACKNOWLEDGEMENTS \end{large}\end{center}}

\input econfmacros.tex

\begin{document}
\begin{titlepage}

\vfill
\Title{Neutrino Mass Ordering Studies with PINGU and IceCube/DeepCore }
\vfill
\Author{ Steven Wren }
\Address{\Manchester}
\vfill
\begin{Abstract}
The Precision IceCube Next Generation Upgrade (PINGU) is a proposed extension to the IceCube detector. The design of PINGU would augment the existing 86 strings with an additional 40 with the main goal of determining the neutrino mass ordering (NMO). Preliminary studies of the NMO can start with IceCube/DeepCore, a sub-array of more densely-packed strings in operation since 2011. This detector has a neutrino energy threshold of roughly 10 GeV and allows for high-statistics datasets of atmospheric neutrinos to be collected. This data provides a unique opportunity to better understand the systematic effects involved in making the NMO measurement by comparing the simulation studies to real data. These proceedings will present the current status of these studies in Monte Carlo simulations with projected DeepCore sensitivity for the NMO. 
\end{Abstract}
\vfill
\begin{Presented}
NuPhys2015, Prospects in Neutrino Physics
Barbican Centre, London, UK,  December 16--18, 2015
\end{Presented}
\vfill
\end{titlepage}
\def\thefootnote{\fnsymbol{footnote}}
\setcounter{footnote}{0}

\section{Introduction}

\noindent The IceCube Neutrino Observatory is the world's largest neutrino detector, instrumenting over \(1\,\unit{km}^3\) of the South Pole ice. Consisting of 5160 optical sensors arranged vertically in 86 `strings' separated by \(\sim120\mathrm{m}\), it has been designed to measure the highest energy neutrinos ever observed. Since 2011, a sub-array of more densely-packed strings, known as DeepCore, has been in operation. This instruments the deepest and most transparent ice, to give an energy threshold of \(\sim\) 10 GeV, allowing high-statistics datasets of atmospheric neutrinos to be collected. It is from this that the first statistically significant observation of the oscillations of neutrinos with energies above 20 GeV was made \cite{bib:DeepCoreOscillations}. Since this publication the results have only improved with more sophisticated reconstruction techniques and a better understanding of this unique medium. 
\\
\indent The planned upgrade for the lower energy neutrino physics at the South Pole is the Precision IceCube Next Generation Upgrade (PINGU), a further 40 strings separated by \(\sim20\mathrm{m}\). This will allow the energy threshold to go as low as 1 GeV while also collecting an order of magnitude more statistics. The main goal of this will be to determine the neutrino mass ordering (NMO). However, studies on the NMO can begin now with the existing DeepCore dataset. Comparing the actual data with the results of Monte Carlo simulations will give a better understanding of the systematics involved in such a measurement and provide a solid framework from which to move on towards PINGU. These proceedings will present the current status of these studies in Monte Carlo simulations with projected DeepCore sensitivity for the NMO.

\section{The Neutrino Mass Ordering}

\begin{figure}[htb]
\centering
\includegraphics[width=11cm]{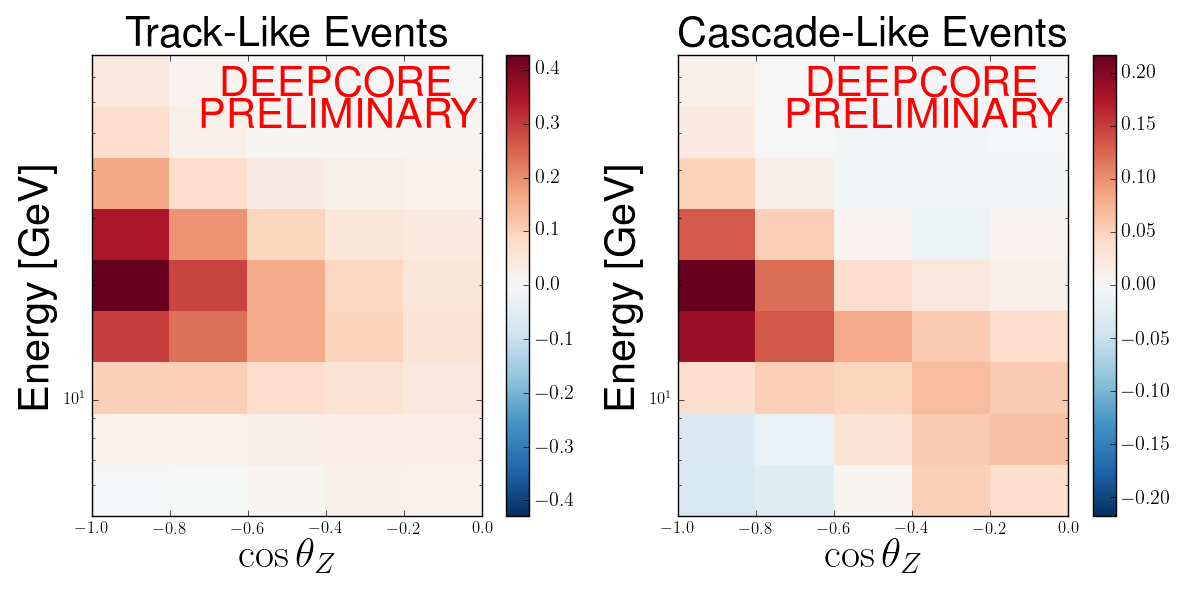}
\vspace{-15pt}
\caption{A visualisation of the NMO asymmetry in the DeepCore data, assuming the neutrino oscillation parameters of reference \cite{bib:NuFit}.}
\label{fig:DeepCoreNMOSignal}
\end{figure}

\noindent Within the standard neutrino oscillation paradigm, experiments make measurements of both the mixing angles, \(\theta_{ij}\), and the squared-mass splittings, \(\Delta m^2_{ij}\), between 3 neutrino mass states, \(i,j = 1,2,3\). The former governs the flavour composition of each of these states and the latter determines the frequency of the oscillations in \(L/E\) space, where \(L\) is the distance the neutrinos have travelled and \(E\) are their energies. However, such measurements do not answer the question of the ordering of these states, i.e. they do not give the sign of \(\Delta m^2_{ij}\). Data from solar neutrinos collected by SNO was found to be consistent with \(\Delta m^2_{21}>0\) \cite{bib:SNO}, but the question of the sign of \(\Delta m^2_{31}\) remains open. The two possibilities, \(\Delta m^2_{31}>0\) and \(\Delta m^2_{31}<0\), are referred to as the normal and inverted ordering (NO and IO) respectively. The current global fit to all neutrino data favours the inverted ordering at a level \(\sim1\sigma\) \cite{bib:NuFit}.
\\
\indent These proceedings will focus on work done towards making the neutrino mass ordering measurement with a dataset of atmospheric neutrinos collected at the South Pole by the DeepCore detector. One can see hierarchy-dependent effects in such datasets due to a combination of matter effects and parametric resonances when the neutrinos travel through the Earth. A visualisation of this signal, following the convention of reference \cite{bib:Akhmedov}, can be seen in figure \ref{fig:DeepCoreNMOSignal}. The metric shown is the difference between the number of events expected for each mass ordering, divided by the square root of the number of events expected for the normal mass ordering; effectively, it is the statistical-only significance of the mass hierarchy signal, in units of \(\sigma\), in each bin. This both highlights the regions in \(\left(E,\cos\theta\right)\) space where an NMO signal is expected to appear and gives a measure of the magnitude of the effect. Here, the events are separated into `track-like' and `cascade-like' classifications. The former is composed ideally of \(\nu_{\mu}\) CC events and everything else falls in to the latter. This separation is done since the directional resolution on cascade-like events is generally poorer, and so one can maximise the signal by fitting these regions separately.

\section{Results}

\begin{figure}[htb]
\centering
\includegraphics[width=7.5cm]{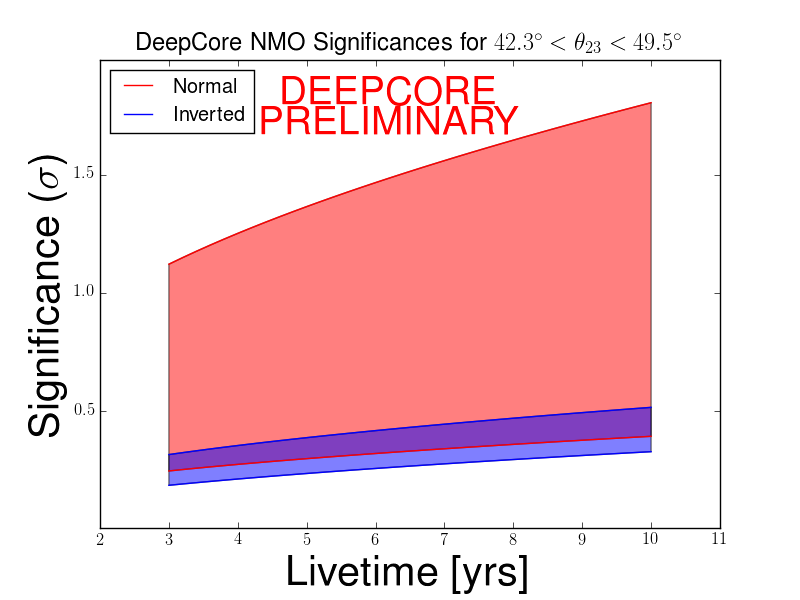}
\includegraphics[width=7.5cm]{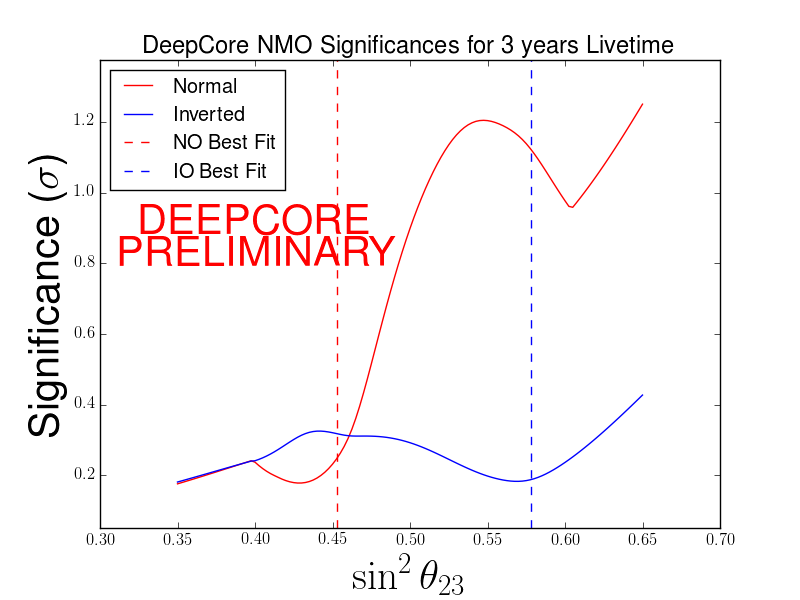}
\vspace{-25pt}
\caption{The expected DeepCore median NMO significances as a function of detector livetime (left) and the true value of \(\sin^2\theta_{23}\) (right). The livetime plot shows the range of significances for \(42.3^{\circ}<\theta_{23}<49.5^{\circ}\) and the \(\sin^2\theta_{23}\) plot shows the best fit points for both the NO and IO from reference \cite{bib:NuFit}.}
\label{fig:DeepCoreNMOSignificances}
\end{figure}

\noindent The expected sensitivity of the DeepCore detector to the NMO is shown in figure \ref{fig:DeepCoreNMOSignificances}. Here, a fast \(\Delta\chi^2\)-based analysis has been performed as in reference \cite{bib:Asimov}. This has been evaluated as both a function of detector livetime and the true value of \(\sin^2\theta_{23}\). For the latter, certain regions of the parameter space see a decrease in the significance. In general, one would expect the NMO signal to grow when one moves to the second octant due to the presence of more \(\nu_{\mu}\) in mass state 3. However, due to degeneracies between this and the true ordering, the signal is obscured. As is evident from the results of reference \cite{bib:NuFit}, the two regimes of \(\left(\mathrm{NO},\theta_{23}<45^{\circ}\right)\) and \(\left(\mathrm{IO},\theta_{23}>45^{\circ}\right)\) are degenerate, and so the NMO significance in the latter is greatly reduced. The case of \(\left(\mathrm{NO},\theta_{23}>45^{\circ}\right)\) is unique in that, broadly speaking, it does not suffer from these strong degeneracies.  Thus, the significances grow as expected. Here, the 10-year results have the potential to exceed \(1.5\sigma\). However, this shows that the biggest obstacle in determining the NMO is the inability to distinguish the octant of \(\theta_{23}\). Nonetheless, the variation of the NMO signal with \(\sin^2\theta_{23}\) leads to significance ranges of \(0.2\sigma\mathrm{\--}0.4\sigma\) and \(0.2\sigma\mathrm{\--}1.2\sigma\) for IO and NO respectively for 3 years of detector exposure. This is approximately how much data has already been collected by the DeepCore detector.

\section{Including Priors on \(\Delta m^2_{31}\) and \(\theta_{23}\)}

\begin{figure}[htb]
\centering
\includegraphics[width=7.5cm]{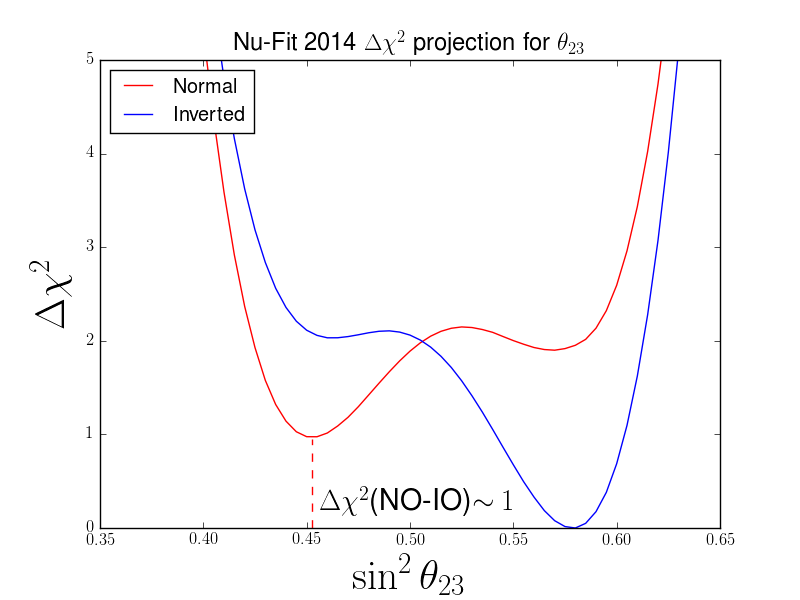}
\includegraphics[width=7.5cm]{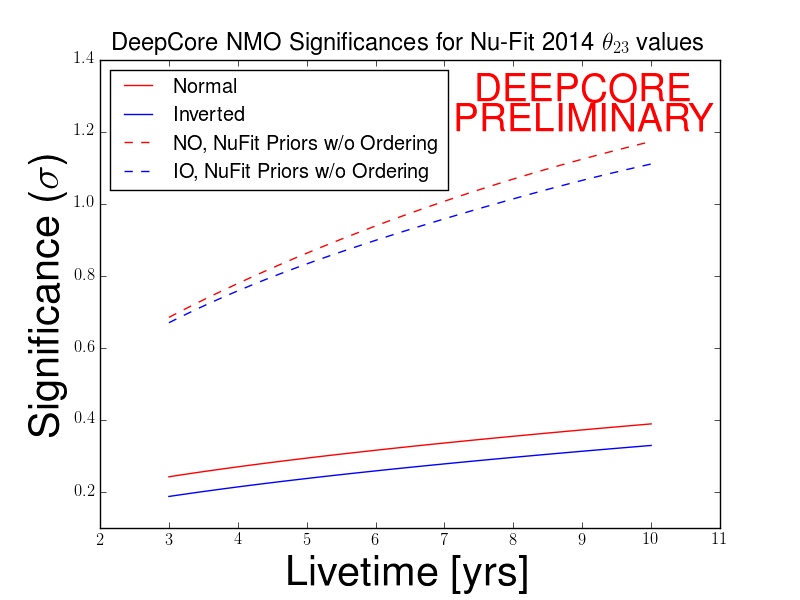}
\vspace{-25pt}
\caption{The expected DeepCore median NMO significances as a function of detector livetime both with (dashed) and without (solid) the priors on \(\theta_{23}\) (shown on the left, from reference \cite{bib:NuFit}) and \(\Delta m^2_{31}\).}
\label{fig:NuFitPriorsIncluded}
\end{figure}

\noindent Since the uncertainty on \(\theta_{23}\) is the largest reason for being unable to determine the NMO, one stands to gain considerably in the measurement by utilising the current global knowledge on this parameter. In this analysis, the \(\chi^2\) surfaces for \(\theta_{23}\) reported in reference \cite{bib:NuFit} have been used as priors. However, using them directly also brings a prior on the ordering itself, since they have a \(\Delta\chi^2\) of \(\sim\) 1 between the two ordering hypotheses, shown on the left in figure \ref{fig:NuFitPriorsIncluded}. This difference in the minima is subtracted before they are used to remove this unwanted effect. A gaussian prior for \(\Delta m^2_{31}\) has also been included based on the \(1\sigma\) error bands in reference \cite{bib:NuFit}, but this is expected to have a significantly smaller effect.
\\
\indent The effect of including these priors as a function of detector livetime is shown on the right in figure \ref{fig:NuFitPriorsIncluded} assuming the fiducial oscillation model of reference \cite{bib:NuFit} only. This leads to a marked increase in sensitivity, with the 3-year results growing from \(\sim\) \(0.2\sigma\) to \(\sim\) \(0.7\sigma\) for both orderings. If this is to be used in the final analysis then the treatment of the remaining systematics will have to be more carefully assessed, as they would no longer be essentially negligible. 

\section{Conclusion and Outlook}

\noindent A \(\Delta\chi^2\)-based analysis to assess the NMO sensitivity of Monte Carlo equivalent to the currently available DeepCore data has been presented. While the predicted significances are low, these studies will assist in developing the analysis with a view towards the IceCube upgrade known as PINGU. 

\Acknowledgements
I am grateful to the IceCube collaboration for all of their invaluable help in performing this work and to my supervisors, Justin Evans and Stefan S{\"o}ldner-Rembold for giving me the chance to work on such an interesting project!

\end{document}

%% file: econfmacros.tex
\def\beq{\begin{equation}}
\def\eeq#1{\label{#1}\end{equation}}
\def\eeqn{\end{equation}}

\def\beqa{\begin{eqnarray}}
\def\eeqa#1{\label{#1}\end{eqnarray}}
\def\eeqan{\end{eqnarray}}

\let\bar=\overbar

\def\Dslash{\not{\hbox{\kern-4pt $D$}}}
\def\dslash{\not{\hbox{\kern-2pt $\del$}}}

\def\msb{{\bar{\ssstyle M \kern -1pt S}}}

